\documentclass[amsmath,notitlepage,amssymb,aps,prl,groupedaddress,twocolumn,footinbib]{revtex4-1}
\usepackage{graphicx,subfigure,epsfig}
\usepackage[pdftex,breaklinks,bookmarks=false,colorlinks,linkcolor=blue,citecolor=blue,urlcolor=blue]{hyperref}
\usepackage{dcolumn}
\usepackage{amssymb}
\usepackage{times}
\usepackage{amsmath}
\usepackage{amsfonts}
\usepackage{mathrsfs}
\usepackage{setspace}
\usepackage{latexsym}
\usepackage{bbm}
\usepackage{float}
\usepackage{flafter}
\usepackage{bm}
\usepackage{epstopdf}
\usepackage{color}
\usepackage{multirow}
\usepackage{physics} 
\usepackage[export]{adjustbox}
\DeclareMathOperator{\sgn}{sgn}

\usepackage[thinc]{esdiff} 

\usepackage{soul} 

\def \be {\begin{eqnarray}}
\def \ee {\end{eqnarray}}

\newcommand{\bk}{{\bf k}}

\newcommand{\bp}{{\bf p}}

\newcommand{\bv}{{\bf v}}

\begin{document}

\title{Probing Fermi sea topology by Andreev state transport}

\author{Pok Man Tam}
\author{Charles L. Kane}
\affiliation{Department of Physics and Astronomy, University of Pennsylvania, Philadelphia, PA 19104, USA}

\begin{abstract}
We show that the topology of the Fermi sea of a two-dimensional electron gas (2DEG) is reflected in the ballistic Landauer transport \textit{along} a long and narrow Josephson $\pi$-junction that proximitizes the 2DEG. The low-energy Andreev states bound to the junction are shown to exhibit a dispersion that is sensitive to the Euler characteristic of the Fermi sea ($\chi_F$). We highlight two important relations: one connects the electron/hole nature of Andreev states to the convex/concave nature of Fermi surface critical points, and one relates these critical points to $\chi_F$. We then argue that the transport of Andreev states leads to a quantized conductance that probes $\chi_F$. An experiment is proposed to measure this effect, from which we predict an $I$-$V$ characteristic that not only captures the topology of Fermi sea in metals, but also resembles the rectification effect in diodes. Finally, we evaluate the feasibility of measuring this quantized response in graphene, InAs and HgTe 2DEGs.
\end{abstract}
\maketitle

\noindent {\color{blue}\emph{Introduction.}} Topological classification of quantum matter has led to discoveries of quantized responses that remain robust under smooth deformation of the physical system \cite{HasanKane2010, QiZhang2011, Chiu2016, Wen2017}. Paradigmatic examples include the one-dimensional (1D) Su-Schrieffer-Heeger chain characterized by the electric polarization \cite{SSH1980, Vanderbilt1993}, and the two-dimensional (2D) integer quantum Hall effect (IQHE) characterized by the Hall conductance \cite{Klitzing1980, TKNN1982}. In these examples, the quantization is associated with the twisting of wavefunction across the Brillouin zone. In metals, there is \textit{another} type of topology associated with the structure of ``holes" in the Fermi sea, as characterized by the Euler characteristic $\chi_F$. For instance, doping graphene above/below charge-neutrality creates two electron/hole-like Fermi pockets, giving $\chi_F = \pm 2$. As every metal can be assigned an Euler number from the topology of its ground state, a fundamental question arises: \textit{does a metal exhibit any quantized response for its Euler number?}

\begin{figure}[b!]
   \includegraphics[width=\columnwidth ]{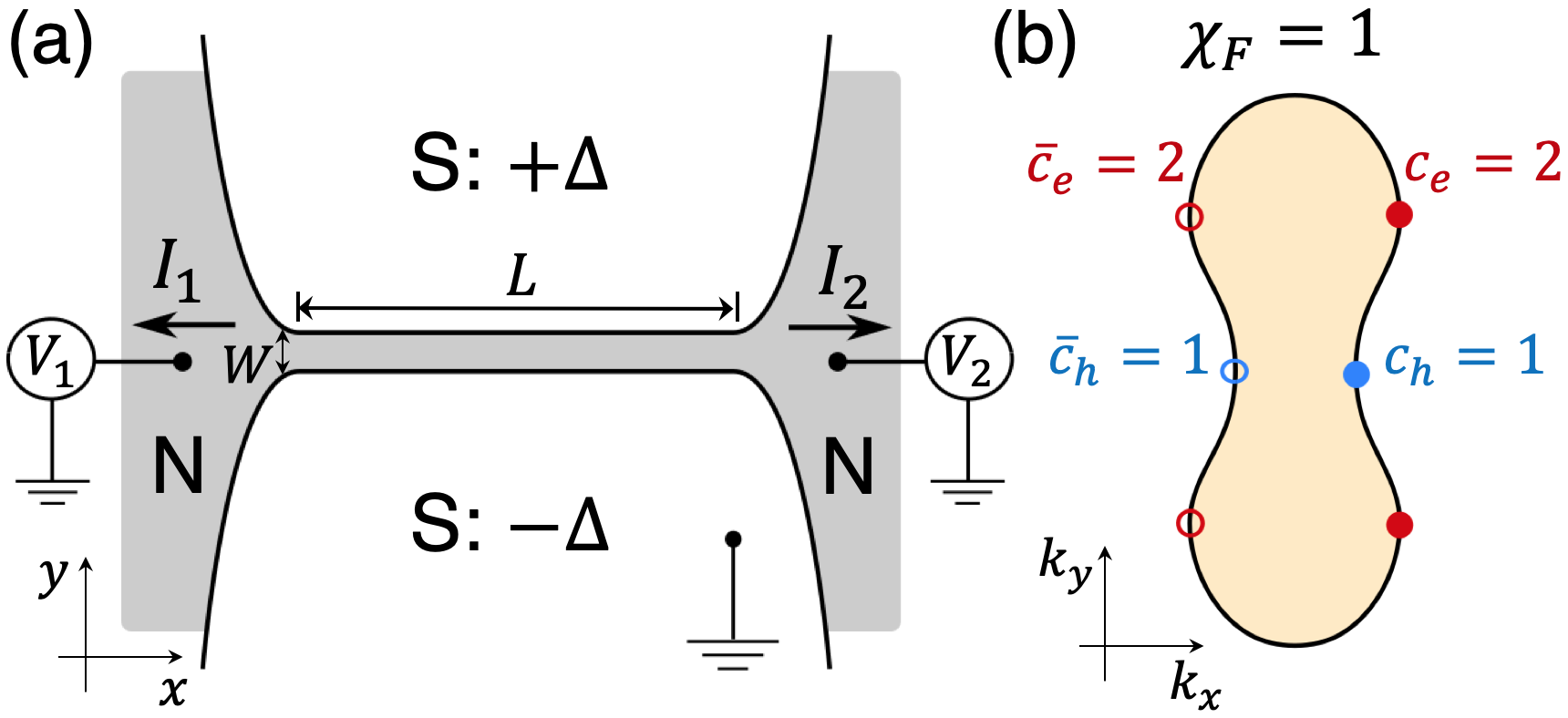}\centering
  \caption{\small{(a) Proposed setup for measuring $\chi_F$ of a 2DEG. Current $I_2$, with $V_2=0$ and $\abs{eV_1}<\abs{\Delta}$, is determined by the Andreev states dispersing along the junction, and encode information about Fermi surface critical points. (b) An electron-pocket Fermi sea, with convex/concave Fermi surface critical points marked in red/blue. Filled/open circles indicate ${\bf v} \parallel \pm\hat{x}$.}}
  \label{figure:setupandFS}
\end{figure}

In 1D, the above question is answered by the Landauer formula \cite{Landauer1957}: for a ballistic conductor with ideal contacts, the linear conductance is $2e^2/h$ times the number of occupied spin-degenerate bands, which is precisely $\chi_F$ in 1D \cite{Betti}. While this quantization is not as robust as the Chern number in the IQHE, it has nevertheless been observed in quantum point contacts \cite{vanWees1988}, semiconductor nanowires \cite{Honda1995, vanWeperen2013}, and carbon nanotubes \cite{Frank1998}. Such successes motivate us to explore higher-dimensional generalizations. Recently, frequency-dependent non-linear response \cite{Kane2022}, equal-time density correlations and multipartite entanglement \cite{Tam2022}, have been proposed to probe $\chi_F$ of a higher-dimensional Fermi sea. The feasibility of measuring a quantized non-linear response in an ultracold atomic gas has been further investigated in Refs. \cite{Yang2022, Zhang2022}. 

In this Letter, we introduce a method to probe $\chi_F$ in 2D metals through nonlocal transport \textit{along} a planar Josephson $\pi$-junction with normal leads, as depicted in Fig. \ref{figure:setupandFS}(a). Similar setups were previously considered for studying topological superconductivity \cite{FuKane2008, Wieder2014, Hell2017, Pientka2017, Akhmerov2018, Fornieri2019, Ren2019, Banerjee2022_1}, while nonlocal transport has emerged as a tool to differentiate topological and trivial phases \cite{Banerjee2022_2, BanerjeeNote}. We show that, in the ballistic limit, biasing the voltage in lead 1 ($V_1$) and measuring the current flowing into lead 2 ($I_2$) results in a quantized two-terminal conductance:
\begin{equation}\label{eq:keyresult}
    G_{21} \equiv \diff{I_2}{V_1} = \frac{2e^2}{h}\big[c_e  \theta(-V_1)+c_h \theta(V_1)\big],
\end{equation}
where $\theta(x)$ is the unit step function. Here $c_e$ and $c_h$ are non-negative integers counting the respective number of convex and concave critical points on the Fermi surface, with velocity ${\bf v} \parallel +\hat{x}$, as illustrated in Fig. \ref{figure:setupandFS}(b). While $c_{e,h}$ depend on the \textit{geometry} of the Fermi surface, as well as the relative orientation of the junction, their difference is only sensitive to the \textit{topology} of the Fermi sea:
\begin{equation}\label{surface_criticalpt}
\chi_F = c_e - c_h.
\end{equation}
Eqs. (\ref{eq:keyresult}) and (\ref{surface_criticalpt}) predict a topological rectification effect: an alternating voltage $V_1(t) = V \sin \omega t$ produces a direct current $\langle I_2 \rangle = -2\chi_Fe^2\abs{V}/\pi h$. Below, we first explain Eq. (\ref{surface_criticalpt}). We then establish a relation between the convex/concave nature of critical points and the particle/hole nature of Andreev states, which disperse along the junction and contribute to $G_{21}$.

\noindent {\color{blue}\emph{Euler characteristic.}} The Euler characteristic $\chi$ was first devised for classifying polyhedra \cite{Euler1758}. For a connected and orientable 2D manifold, $\chi$ is related to the genus $g$ and the number of boundaries $b$ \cite{Nakahara1990, Dieck2008}:
$\chi = 2-2g-b$. Here we classify metals according to the Euler characteristic $\chi_F$ of the \textit{Fermi sea} \cite{FermiCompact}. For instance, an electron pocket (being topologically equivalent to a disk) has $\chi_F=1$, while a hole pocket has $\chi_F=-1$ (as the Brillouin zone is equivalent to a $g=1$ torus). Generally, for 2D metals, $\chi_F$ is the number of electron-like Fermi surfaces \textit{minus} the number of hole-like Fermi surfaces.

Morse theory provides a connection between $\chi_F$ and the energy dispersion $E(\bk)$, which can be viewed as a Morse function. We have $\chi_F = \sum_{\gamma} (-1)^\gamma C_\gamma$, where $C_\gamma$ counts the number of critical points \textit{inside} the Fermi sea with Morse index $\gamma$ \cite{Milnor1963, Nash1988}. A critical point where $\bv = \nabla_{\bk} E(\bk)/\hbar = 0$ can be classified as a minimum, saddle or maximum point, with index  $\gamma=0,1,2,$ respectively. $\gamma$ is related to the Hessian $\mathbb{H}$ of $E(\bk)$: $(-1)^\gamma=\sgn[\det \mathbb{H}]$. It follows that
\begin{equation}\label{sea_criticalpt}
\chi_F = \int d^2 k f_0(\bk) [\frac{\partial \theta(v_x)}{\partial k_x} \frac{\partial \theta(v_y)}{\partial k_y} - \frac{\partial \theta(v_y)}{\partial k_x}\frac{\partial \theta(v_x)}{\partial k_y}],
\end{equation}
with $f_0(\bk) \equiv \theta(E_F - E(\bk))$ defining the Fermi sea.

It is illuminating to convert the integral in Eq. (\ref{sea_criticalpt}) into a form that only receives contribution on the Fermi surface. This is achieved by adding zero to the integrand of the form $\theta(v_x) \partial_{k_x}{\theta(v_y)}\partial_{k_y}{f_0}$ \cite{Kane2022, addzero}, and integrate by parts to obtain
\begin{equation}\label{eq:FSintegral}
\chi_F = -\int d^2k \frac{\partial f_0(\bk)}{\partial k_x} \frac{\partial \theta(v_y)}{\partial k_y} \theta(v_x).
\end{equation}
The first term $\partial f_0 /\partial k_x$ isolates the Fermi surface, while $\theta(v_x) \partial \theta(v_y)/\partial k_y$ isolates the critical points \textit{on the Fermi surface}, where $v_y=0$ and $v_x > 0$. These points are called \textit{convex} (if $\partial_{k_y} v_y > 0$) or \textit{concave} (if $\partial_{k_y} v_y < 0$), whose local neighbourhood resembles an \textit{electron}-like or a \textit{hole}-like Fermi surface respectively. Denoting the number of convex/concave critical points as $c_{e/h}$, we arrive at Eq. (\ref{surface_criticalpt}). Alternatively, $\chi_F = \bar{c}_e -\bar{c}_h$, where $\bar{c}_{e,h}$ count critical points with $v_y =0 $ and $v_x < 0$. Time-reversal symmetry requires $c_{e,h}=\bar{c}_{e,h}$. 

We will first neglect spin-orbit interactions, so $c_{e,h}$, $\bar{c}_{e,h}$ and $\chi_F$ are defined for the spin-degenerate Fermi sea. In the end, and in the supplementary \cite{supp}, we address the effect of Rashba spin-orbit coupling (SOC), and argue that our results remain valid.

\noindent {\color{blue}\emph{Critical points and Andreev states.}} Let us now relate Fermi surface critical points to properties of Andreev bound states (ABS) formed at the SNS junction (S denotes an $s$-wave superconductor, N denotes the normal metal of interest). We consider a junction geometry shown in Fig. \ref{figure:setupandFS}(a), with $L \gg \xi \gg W$. Here, $L$ is the length of the junction, $\xi$ the superconducting coherence length and $W$ the distance between two SN interfaces. The narrow-junction minimizes the number of ABSs, and the long-junction suppresses crossed Andreev reflections between leads. We also take the adiabatic limit, in which all potentials vary smoothly on the scale of the Fermi wavelength $k_F^{-1}$. With $k_F\xi \gg 1$, the pairing gap does not alter the topology of the Fermi sea.

Our proposal concerns a $\pi$-junction, across which the superconducting order parameter changes sign. A distinguished feature of $\pi$-junctions is the existence of \textit{zero energy} ABSs \cite{Kulik1969, BvH, Sauls2018}. Intuitively, ABSs are formed by mixing between electron and hole states on the Fermi surface due to Andreev reflection \cite{Andreev1964, *Andreev1966}. Away from Fermi surface critical points, states near the Fermi surface are governed by a \textit{linear} dispersion (in the $y$-direction, or perpendicular to the SN-interfaces). Correspondingly, the Bogoliubov-de Gennes (BdG) equation takes the form of a 1D Dirac equation with a spatially varying mass term \cite{Sauls2018}. The $\pi$ phase-change in the pairing potential implies a ``kink" configuration of the mass term, leading to a Jackiw-Rebbi zero mode localized at the domain wall \cite{Jackiw1976}. 

\begin{figure}[t]
   \includegraphics[width =\columnwidth]{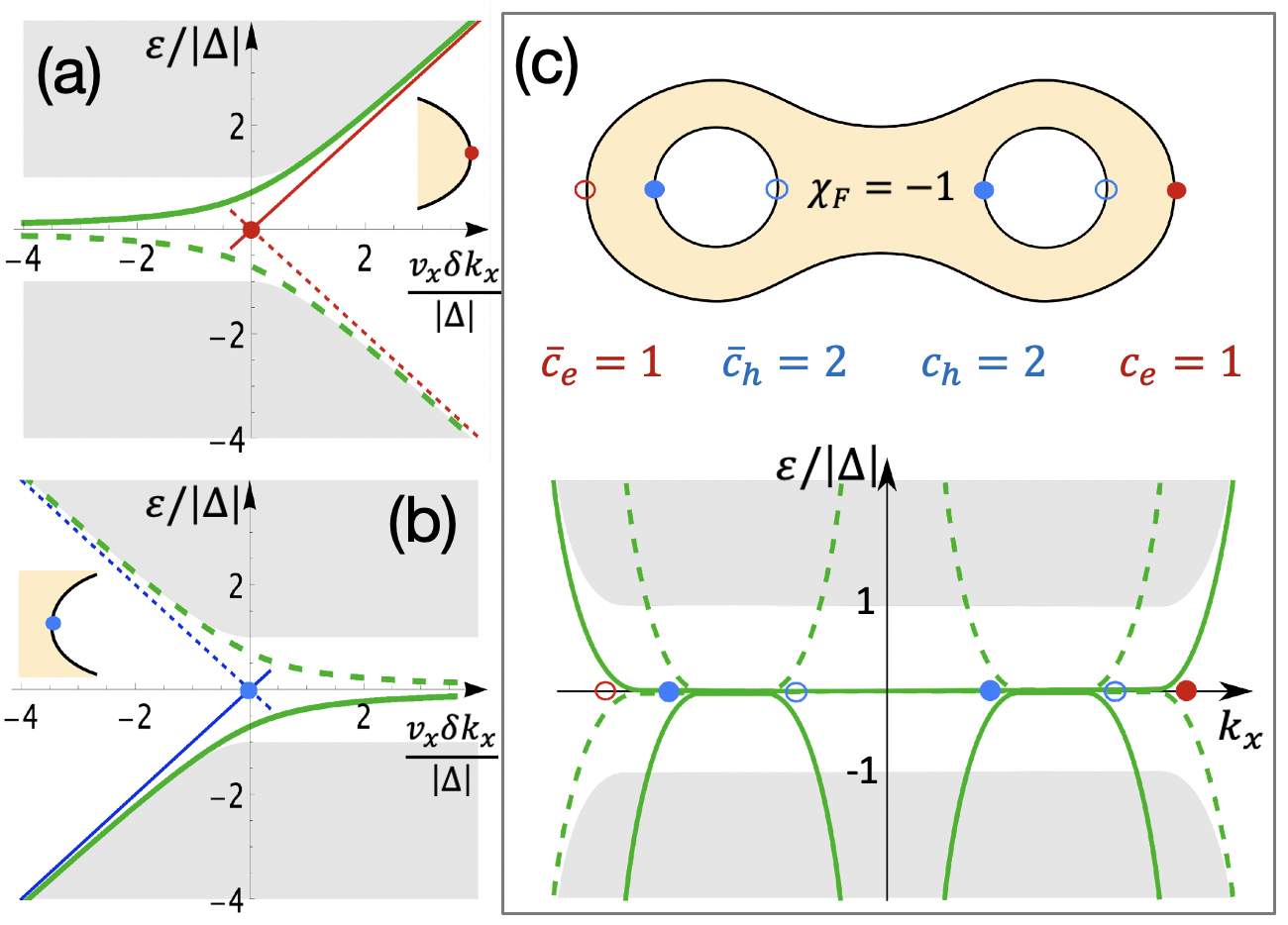}\centering
  \caption{\small{BdG spectrum of an SNS $\pi$-junction, obtained around (a): a convex critical point (i.e. $a>0$, marked in red) and (b): a concave critical point (i.e. $a<0$, marked in blue). Shaded regions represent the continuum of bulk states. The ABS dispersion is plotted in green, using Eq. (\ref{eq:dispersion}), with the solid curve representing the physical state. The solid ray indicates the \textit{normal} band edge. Dashed lines are the particle-hole partners of the solid ones (related by $\tau_x$). (c): Top panel shows a Fermi sea with $\chi_F=-1$; Lower panel shows the schematic BdG spectrum deduced from (a) and (b). There are $c_{e/h}$ right-moving (and $\bar{c}_{e/h}$ left-moving) electron/hole-like ABSs.}}
  \label{fig:dispersion}
\end{figure}

The above picture breaks down close to a Fermi surface critical point, where the dispersion is quadratic. Aside from Andreev reflections, one also needs to consider normal reflections that hybridize electrons at $+k_y$ and $-k_y$ (which coincide at the critical point). This gives rise to a dispersion of the ABS \textit{along} the junction (i.e. the $x$-direction). As shown in Fig. \ref{fig:dispersion}(a, b), on one side of a critical point the ABS is dispersionless (with almost zero energy), while a significant dispersion begins near the critical point, such that on the other side the ABS approaches the bulk band edge. The dispersion implies that the ABS propagates along the junction, and eventually goes into the lead. The process of entering the lead can be modeled in two ways: fixing $W$ while taking the superconducting gap $\abs{\Delta} \rightarrow 0$ (as done below), or fixing $\abs{\Delta}$ while taking $W \rightarrow \infty$ (see the supplementary \cite{supp}). In both formulations, provided the transition is adiabatic, the positive-energy ABS converts into a particle/hole if the associated critical point is convex/concave. Combining this with Eq. (\ref{surface_criticalpt}) allows us to probe $\chi_F$ through transport. Next, we support these claims by examining the dispersion of the ABS obtained from the BdG equation.

\noindent {\color{blue}\emph{Dispersion of ABS.}} Consider a narrow SNS $\pi$-junction in the limit $W \rightarrow 0$. Expanding around a Fermi surface critical point, we obtain the following BdG Hamiltonian:
\begin{equation}\label{BdG}
    H_{\text{BdG}} = (-a \partial^2_y + v_x \delta k_x) \tau_z + \text{sgn}(y) \abs{\Delta} \tau_y.
\end{equation}
Here, $\tau$'s are the Pauli matrices acting on the particle-hole space, and $\abs{\Delta}$ is the magnitude of the $s$-wave pairing gap. In the adiabatic limit we assume translation invariance along the junction, so the momentum $k_x$ is conserved. $\delta k_x$ measures the deviation away from the critical point of interest, where the Fermi velocity is $(v_x,0)$. $a$ is related to the electron's effective mass, with sgn($a$)$=\pm$ for a convex/concave critical point. Note the $W \rightarrow 0$ limit is taken only for convenience, as an analytic solution for the single ABS is available. Finite $W$ is studied in the supplementary \cite{supp}, where a single ABS, with similar features, is found provided $W\lesssim \sqrt{\abs{a}/\abs{\Delta}}$. 

The critical-point model has two symmetries useful for finding the ABS. (1) Chiral symmetry $\Pi=\tau_x$, with $\{H_{\text{BdG}},\Pi \}=0$. $\Pi$ relates states of energy $\pm \varepsilon$ for a fixed $\delta k_x$. (2) Mirror symmetry $\mathcal{M}_y = \tau_z M_y$ (where $M_y$ takes $y \mapsto -y$), with $[H_{\text{BdG}}, \mathcal{M}_y]=0$. Each ABS can be labeled by $\mathcal{M}_y=\pm 1$. As $\{\Pi, \mathcal{M}_y\}=0$, particle-hole partners (related by $\Pi$) acquire opposite mirror eigenvalues. It is sufficient to focus on the $\mathcal{M}_y=1$ sector.

Denote $\Psi(y)$ as the solution of $H_{\text{BdG}} \Psi = \varepsilon \Psi$. As $\Delta$ is uniform away from $y = 0$, $\Psi(y)$ is a linear combination of attenuated waves of the form $e^{-\kappa y}$, with
\begin{equation}\label{eq:kappa}
    a\kappa^2 = v_x \delta k_x \pm i \sqrt{\abs{\Delta}^2-\varepsilon^2}.
\end{equation}
Only two of the four solutions, denoted $\kappa^{\pm}$ (with $\text{Re}(\kappa^{\pm})>0$, $\text{Im}(\kappa^\pm) \gtrless 0$), correspond to bound states for $y>0$, so
\begin{equation}
    \Psi_{y>0} = \gamma^+ \begin{pmatrix}
1 \\ ie^{\text{sgn}(a)i\eta}
\end{pmatrix} e^{-\kappa^+ y} + \gamma^- \begin{pmatrix}
1 \\ ie^{-\text{sgn}(a)i\eta}
\end{pmatrix} e^{-\kappa^- y}.
\end{equation}
The phase $e^{\pm \text{sgn}(a) i\eta} \equiv  \varepsilon/\abs{\Delta} \pm i\text{sgn}(a) \sqrt{1-(\varepsilon/\abs{\Delta})^2}$ characterizes the mixing between particle and hole due to Andreev reflections, while $\gamma^{+}$ and $\gamma^{-}$ account for normal reflections.

Using the mirror symmetry, the ABS wavefunction in the lower plane is $\Psi(y<0) = \mathcal{M}_y \Psi(y>0)$. Matching two boundary conditions required by the quadratic BdG equation, we find
\begin{subequations}
    \begin{align}
        \Psi(0^-) = \Psi(0^+) &:\quad \gamma^-/\gamma^+ = -e^{2i\text{sgn}(a) \eta}, \\
        \partial_y\Psi(0^-)= \partial_y\Psi(0^+) &:\quad \gamma^-/\gamma^+ = -\kappa^+/\kappa^-.
    \end{align}
\end{subequations}
These, together with Eq. (\ref{eq:kappa}), give the dispersion of the ABS:
\begin{equation}\label{eq:dispersion}
    \varepsilon = \pm \Big[\frac{v_x \delta k_x}{2} + \text{sgn}(a) \sqrt{\Big(\frac{v_x\delta k_x}{2}\Big)^2+\frac{\abs{\Delta}^2}{2}}\Big].
\end{equation}
Figure \ref{fig:dispersion} (a) and (b) show a plot of this for $a>0$ and $a<0$ respectively. Right at the Fermi surface critical point ($\delta k_x =0$), we find $\varepsilon = \pm \abs{\Delta}/\sqrt{2}$. Away from it, and on the \textit{inside} of the Fermi pocket, we have $\varepsilon \approx \pm \abs{\Delta}^2/(2v_x\delta k_x)$. Deep inside the Fermi pocket, the ABS thus becomes an approximate zero mode, which is consistent with the familiar result obtained by neglecting normal reflections \cite{Kulik1969, BvH, Sauls2018}. 

On the other side of the critical point (\textit{outside} the Fermi pocket), we obtain $\varepsilon \approx \pm v_x\delta k_x$. Notice that the band edge of the \textit{normal} dispersion, which describes physical states in the normal region (without BdG doubling), is precisely located at $E_{\text{edge}}=v_x \delta k_x$. This leads us to classify two types of ABS: (i) \underline{Electron-like ABS}, which converges to the normal band edge $E_{\text{edge}}$ for $\varepsilon >0$ and appears near a convex critical point ($a>0$); (ii) \underline{Hole-like ABS}, which converges to the normal band edge $E_{\text{edge}}$ for $\varepsilon <0$ and appears near a concave critical point ($a<0$). 

An electron/hole-like ABS inside the junction would convert definitively into an electron/hole when it propagates into the normal leads. This is because for a fixed $\varepsilon$, as $\abs{\Delta} \rightarrow 0$, the corresponding ABS is pushed deep into the regime where its dispersion coincides with either the electron-like or the hole-like $E_\text{edge}$. It thus acquires their electron/hole character. The same conclusion is reached when we model the lead by widening the junction ($W \rightarrow \infty$) \cite{supp}. This is our second key relation: there is an electron/hole-like ABS for each convex/concave Fermi surface critical point. Moreover, each ABS (by which we mean specifically the \textit{physical} state that can adiabatically continue into the normal region) propagates along the junction in the same direction as $v_x$. Figure \ref{fig:dispersion}(c) illustrates this relation for a Fermi sea with $\chi_F=-1$. 

\noindent {\color{blue}\emph{Quantized conductance and topological rectification.}} Now we propose an experiment to extract $\chi_F$ from a two-terminal transport \textit{along} an SNS $\pi$-junction. The setup is depicted in Fig. \ref{figure:setupandFS}(a). Let us fix $V_2=0$, so that no Andreev reflection happens between lead 2 and the superconductor. Then $I_2$ is solely contributed by normal transmissions between leads. Using the Landauer formalism we find \cite{Datta1995, supp},
 \begin{equation}\label{eq:Landauer}
I_2 = \frac{2e}{h}\int_{-\infty}^\infty  T_{21}(E) \; [f(E-eV_1)-f(E)] \; dE.
\end{equation}
The factor of 2 is due to spin (SOC is considered later) and $f(E)=(e^{\beta E}+1)^{-1}$ is the Fermi distribution at temperature $\beta^{-1}$. $T_{21} (E)$ describes transmission from lead 1 to lead 2, which we denote by $T_{21}^{e/h}$ for $E \gtrless 0$. For $|eV_1|< |\Delta|$ the only available channels are the electron/hole-like dispersive ABSs, which derive from the convex/concave critical points. Hence
\begin{equation}\label{eq:transmission}
    T^e_{21}(E) = c_e \,   \theta(E-\delta), \;\; T^h_{21}(E) = c_h \,  \theta(-E-\delta),
\end{equation}
with reflectionless contacts and ballistic transport assumed.

\begin{figure}[t!]
   \includegraphics[width=\columnwidth]{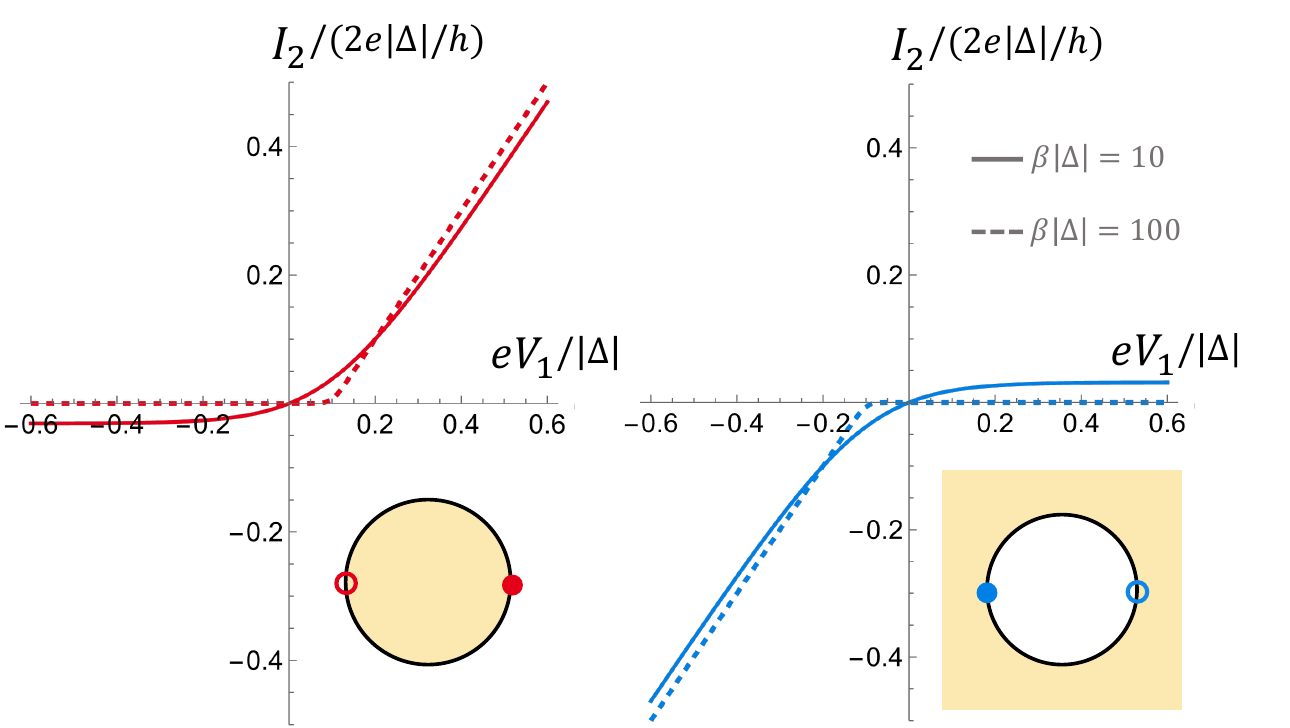}\centering
  \caption{\small{$I_2$-$V_1$ characteristic curves for transport along the junction, revealing the shape of the Fermi sea. In the ohmic regime, the slope (in unit of $2e^2/h$) is quantized to $c_{e/h}$ for $V_1 \lessgtr 0$. Red/blue curve shown here corresponds to an electron/hole-like Fermi pocket with $c_{e/h}=1$ and $c_{h/e}=0$}. Here $\delta=0.1\abs{\Delta}$.}
  \label{fig:finiteT}
\end{figure}

Parameter $\delta$ is introduced to characterize the splitting of ABSs. For a $\pi$-junction, while $\delta = 0$ in the linearized model with exact zero modes, there are realistic reasons for $\delta>0$. For instance, a parabolic normal dispersion $H_0 = {\bf p}^2/2m-E_F$ implies $\delta \approx \abs{\Delta^2/(2E_F)}$. For an ideal graphene SNS junction with $H_0 = v\bp\cdot \mathbf{\sigma}-E_F$, one finds $\delta=0$ \cite{Titov2006}, but an imperfect junction transparency causes $\delta >0$ \cite{Bretheau2017, Park2022}. Different ABSs would also have different splittings in general. Nonetheless, our working assumption is $\delta \ll \abs{\Delta}$, so we include one parameter to model these realistic effects. 

Following Eqs. (\ref{eq:Landauer}, \ref{eq:transmission}), we obtain
\begin{equation}\label{eq:I-V}
    I_2(V_1) = \frac{2e}{h} \big[ c_e \, g(V_1) - c_h \, g(-V_1)\big], 
\end{equation}
with 
\begin{equation}
\begin{split}
    g(V)& \;\;=\;\; \beta^{-1}\ln[(1+e^{\beta(eV-\delta)})/(1+e^{-\beta\delta})] \\
    &\xrightarrow{\beta\rightarrow\infty}\;(eV-\delta)\; \theta(eV-\delta).
\end{split}
\end{equation}
For $k_B T = \delta =0$, the two-terminal conductance $G_{21}\equiv\text{d}I_2/\text{d}V_1$ is quantized in the form of Eq. (\ref{eq:keyresult}). This quantization remains robust provided $\abs{\Delta}>\abs{eV_1}\gg\delta, k_BT$.

Equation (\ref{eq:I-V}) implies a unique $I$-$V$ characteristic for transport along the SNS $\pi$-junction. Figure \ref{fig:finiteT} shows the $I_2$-$V_1$ curves for two simplest geometries of the Fermi sea, which exhibit distinct \textit{unidirectional} behaviors. Different from the Andreev rectifier proposed in Ref. \cite{Akhmerov2018}, the rectification predicted here is intrinsic to the topology of a metal. For generic shapes of the Fermi surface, and for different orientations of the junction, integers $c_{e/h}$ could vary and lead to different slopes of the $I$-$V$ curve in the backward/forward-biased regime. Nevertheless, $\chi_F=c_e-c_h$ is a robust topological quantity, insensitive to either the Fermi surface geometry or the junction orientation.

\noindent {\color{blue}\emph{Discussion.}} While SOC is ignored in the above analysis, its effect can be easily incorporated. Let us include the Rashba term $\lambda_R({\bf k} \times {\bf \sigma})\cdot \hat{z}$ into the Hamiltonian, with $\sigma$ being the spin Pauli matrix. Turning on $\lambda_R>0$ splits the spin-degenerate Fermi surfaces by shrinking/enlarging the one with $\sgn[({\bf k} \times {\bf \sigma})\cdot \hat{z}]=\pm 1$. As before, we begin our analysis deep inside both Fermi pockets, so that we can linearize around the Fermi surface. The BdG Hamiltonian for the $\pi$-junction is then
\begin{equation}
\begin{split}
    H(k_x) = &\sgn(y)\abs{\Delta}\tau_y+\{-i\mu_z v_y \partial_y+ \\
    &\lambda_R[k_x\sigma_y-(\mu_z k^F_y-i\partial_y)\sigma_x]\}\tau_z, 
\end{split}
\end{equation}
where $\mu_zv_y=\pm v_y$ is the Fermi velocity for the up/down-mover around the Fermi point at $\pm k^F_y$. For $k_x=0$, there are two pairs of Jackiw-Rebbi zero modes with $\mu_z\tau_x=+1$. For generic $k_x$, $H(k_x)$ belongs to the symmetry class CII \cite{Teo2010, classCII}, with the chiral symmetry implemented by $\Pi'=\mu_z\tau_x$. Hence the zero modes, which all share the same $\Pi'$-eigenvalue, cannot hybridize for generic $k_x$. Nonetheless, the above argument breaks down when $k_x$ approaches Fermi surface critical points, around which ABSs would begin dispersing due to normal reflections, and eventually merge with the bulk continuum outside the Fermi pockets. This is confirmed by obtaining the full dispersion of ABSs in the Rashba model \cite{supp}. We shall thus replace $2\chi_F$ in previous discussions by $\chi^{\text R}_F+\chi^{\text L}_F$, i.e. the sum of Euler characteristics for the two Fermi seas with right(R)- and left(L)-handed spin-momentum locking. Similarly, $2c_{e,h} \rightarrow c^{\text R}_{e,h}+c^{\text L}_{e,h}$. As such, our quantized-transport formulas remain valid in the presence of Rashba SOC.

Finally, we suggest specific avenues for realizing our proposal. One platform is the InAs or HgTe 2DEG, which has been fabricated into planar junctions for studying topological superconductivity \cite{Fornieri2019, Ren2019, Banerjee2022_1}. Measurements of the nonlocal conductance ($G_{21}$) have been reported recently \cite{Banerjee2022_2, BanerjeeNote}. Another avenue, where SOC is much weaker, is the graphene-based Josephson junction \cite{Titov2006}, where tunneling spectroscopy has been performed to measure the spectrum of ABSs \cite{Bretheau2017, Park2022}. The ability to tune the chemical potential through charge-neutrality, where $\chi_F$ (per spin) changes between $\pm 2$, allows one to demonstrate an abrupt jump in conductance. To observe a robust quantization, a \textit{ballistic}, \textit{long} and \textit{narrow} junction ($ \ell \gg L \gg \xi \gg W$) is required, where $\ell$ is the mean free path. While narrow planar junctions are available \cite{Fornieri2019,Ren2019, Bretheau2017}, the challenge is posed by a short mean free path ($\ell \sim \xi$) in such systems. Away from the ballistic limit, while the quantization is no longer exact, $G_{21}$ is still expected to transit drastically as the topology of Fermi sea changes. In summary, our work calls for near-term experimental effort to look for quantized transport along a planar SNS $\pi$-junction, which probes the intrinsic topology of 2D metals.

\begin{acknowledgments}
\noindent {\color{blue}\emph{Acknowledgments.}} We thank Ady Stern for helpful discussions.   This work was supported by a Simons Investigator Grant to C.L.K. from the Simons Foundation.
\end{acknowledgments}

\bibliographystyle{apsrev4-1.bst}

\begin{thebibliography}{48}
\bibitem [{Bet()}]{Betti}%
  \BibitemOpen
  \href@noop {} {}\bibinfo {note} {Modern definition of $\chi$ involves the
  alternating sum of Betti numbers: $\chi = b_0-b_1+b_2...$, where $b_\ell$
  counts the topological classes of $\ell$-cycles on the manifold. $b_0$ counts
  the number of connected components. For a 1D Fermi sea, $b_0$ thus counts the
  number of occupied bands. In the main text, we discuss more detailed ways to
  calculate $\chi$ in 2D.}\BibitemShut {Stop}%
\bibitem [{Fer()}]{FermiCompact}%
  \BibitemOpen
  \href@noop {} {}\bibinfo {note} {Periodicity of the Brillouin zone ensures
  that the Fermi sea is a compact manifold, possibly with multiple connected
  components.}\BibitemShut {Stop}%
\bibitem [{add()}]{addzero}%
  \BibitemOpen
  \href@noop {} {}\bibinfo {note} {This term is clearly zero, as $\partial
  f_0/\partial{k_y} = v_y \partial f/\partial E$, while
  $\partial{\theta(v_y)}/\partial{k_x}$ fixes $v_y=0$.}\BibitemShut {Stop}%
\bibitem [{cla()}]{classCII}%
  \BibitemOpen
  \href@noop {} {}\bibinfo {note} {$H(k_x)$ has two anti-unitary symmetries,
  $\Theta'=\mu_y\sigma_z\tau_z\mathcal{K}$ and
  $\Xi'=\mu_x\sigma_z\tau_y\mathcal{K}$, with
  $\Theta'^2=\Xi'^2=-1$.}\BibitemShut {Stop}%
\end{thebibliography}

\newpage
\widetext
\begin{center}
\textbf{\large Supplementary Materials for ``Probing Fermi sea topology by Andreev state transport"}\\
\vspace{0.5cm}
\text{Pok Man Tam and Charles Kane}
\end{center}
\onecolumngrid
\setcounter{secnumdepth}{2}
\renewcommand{\theequation}{\thesubsection.\arabic{equation}}
\renewcommand{\theHequation}{\theHsubsection.\arabic{equation}}
\renewcommand{\thefigure}{\thesubsection.\arabic{figure}}    

We present three supplementary sections to provide further details to our discussion in the main text. In Sec. \ref{supp:finiteW}, we model a normal lead as a wide SNS junction, from which we verify our claim about the fate of an Andreev bound state (ABS) as it moves from a narrow junction into the lead. In Sec. \ref{supp:LB}, we apply the Landauer formalism to an SNS junction and derive Eq. (10) in the main text, which allows us to relate the transport properties along the junction to the topology of Fermi sea. In Sec. \ref{supp:SOC}, we solve for the dispersive ABSs in the presence of Rashba spin-orbit coupling (SOC). Our solution confirms that there is one branch of electron/hole-like ABS for each electron/hole-like Rashba-split Fermi pocket, which disperses drastically around a Fermi surface critical point.

\subsection{Modeling the lead: Andreev states in a wide junction}\label{supp:finiteW}
\setcounter{equation}{0}
\setcounter{figure}{0}   
In the main text, we found an analytic expression for the dispersion of ABS in an extremely narrow ($W=0$) SNS $\pi$-junction, and deduced the fate of the ABS excitation as it moves from the junction into the normal lead: the ABS turns into an electron/hole excitation if the normal dispersion is an electron/hole-like band. There the process of propagating from the junction into the lead is modeled by adiabatically taking the pairing gap $\abs{\Delta} \rightarrow 0$. Here we provide an alternative picture by modeling the lead as a \textit{wide} junction with $W \gg \xi$ ($\xi$ is the superconducting coherence length). Propagating from the narrow junction into the normal lead then corresponds to following an ABS as $W$ is increased adiabatically (as represented in Fig. 1(a) of the main text). These two pictures ($\abs{\Delta} \rightarrow 0$ and $W \rightarrow \infty$) are equivalent, as demonstrated below.

As in the main text, we focus on the neighbourhood of a Fermi surface critical point, and consider the following Bogoliubov-de Gennes (BdG) Hamiltonian:
\begin{equation}\label{supp:eq:fW_Ham}
    H_{\text{BdG}} = (-a \partial^2_y +v_x\delta k_x) \tau_z +[\theta(y-W/2)-\theta(-y-W/2)]\cdot \abs{\Delta} \tau_y
\end{equation}
This is similar to Eq. (5) in the main text, so we are not explaining again the meaning of parameters introduced before. A major difference is that we now have a normal region of width $W>0$, sandwiched between two superconductors with opposite pairing phases. Just like Eq. (5), this model has the same chiral symmetry ($\Pi=\tau_x$) and mirror symmetry $(\mathcal{M}_y=\tau_z M_y)$. Hence, it is again sufficient to focus on the $\mathcal{M}_y=1$ sector, and use $\Pi$ to obtain the full spectrum which is particle-hole symmetric. First consider region (I) with $-W/2 < y < W/2$. In this normal region, the ABS at energy $\varepsilon$ should have a wavefunction of the following form,
\begin{equation}
    \Psi_{\text I}(y) = \Gamma_e^+ \begin{pmatrix}
1 \\ 0
\end{pmatrix} e^{ik_e y} + \Gamma_e^- \begin{pmatrix}
1 \\ 0
\end{pmatrix} e^{-ik_e y}+  \Gamma_h^+ \begin{pmatrix}
0 \\ 1
\end{pmatrix} e^{ik_h y}+ \Gamma_h^- \begin{pmatrix}
0 \\ 1
\end{pmatrix} e^{-ik_h y}.
\end{equation}
Here $ak_{e/h}^2=-v_x\delta k_x \pm \varepsilon$, which is required by the BdG equation $H_{\text{BdG}} \Psi = \varepsilon\Psi$. As $\mathcal{M}_y \Psi=\Psi$, we have $\Gamma_e^+=\Gamma_e^-$ and $\Gamma_h^+=-\Gamma_e^-$. We can thus express,
\begin{equation}
    \Psi_{\text I}(y) = \begin{pmatrix}
\Gamma_e \cos{k_e y} \\ \Gamma_h \sin{k_h y}
\end{pmatrix}.
\end{equation}
Next, we consider region (II) with $y>W/2$, which is inside a superconductor. A generic bound-state solution takes the exact same form as Eq. (7) in the main text, i.e.
\begin{equation}
    \Psi_{\text{II}} (y) = \gamma^+ \begin{pmatrix}
1 \\ ie^{\text{sgn}(a)i\eta}
\end{pmatrix} e^{-\kappa^+ (y-W/2)} + \gamma^- \begin{pmatrix}
1 \\ ie^{-\text{sgn}(a)i\eta}
\end{pmatrix} e^{-\kappa^- (y-W/2)}, 
\end{equation}
with $e^{\pm \text{sgn}(a) i\eta} \equiv  \varepsilon/\abs{\Delta} \pm i\text{sgn}(a) \sqrt{1-(\varepsilon/\abs{\Delta})^2}$. Again, $\kappa^{\pm}$ are the two solutions for $a\kappa^2=v_x\delta k_x \pm i\sqrt{\abs{\Delta}^2-\varepsilon^2}$ that satisfy $\Re(\kappa^\pm)>0$ and $\Im(\kappa^\pm)\gtrless0$. 

\begin{figure}[h]
  \includegraphics[width=0.9\columnwidth ]{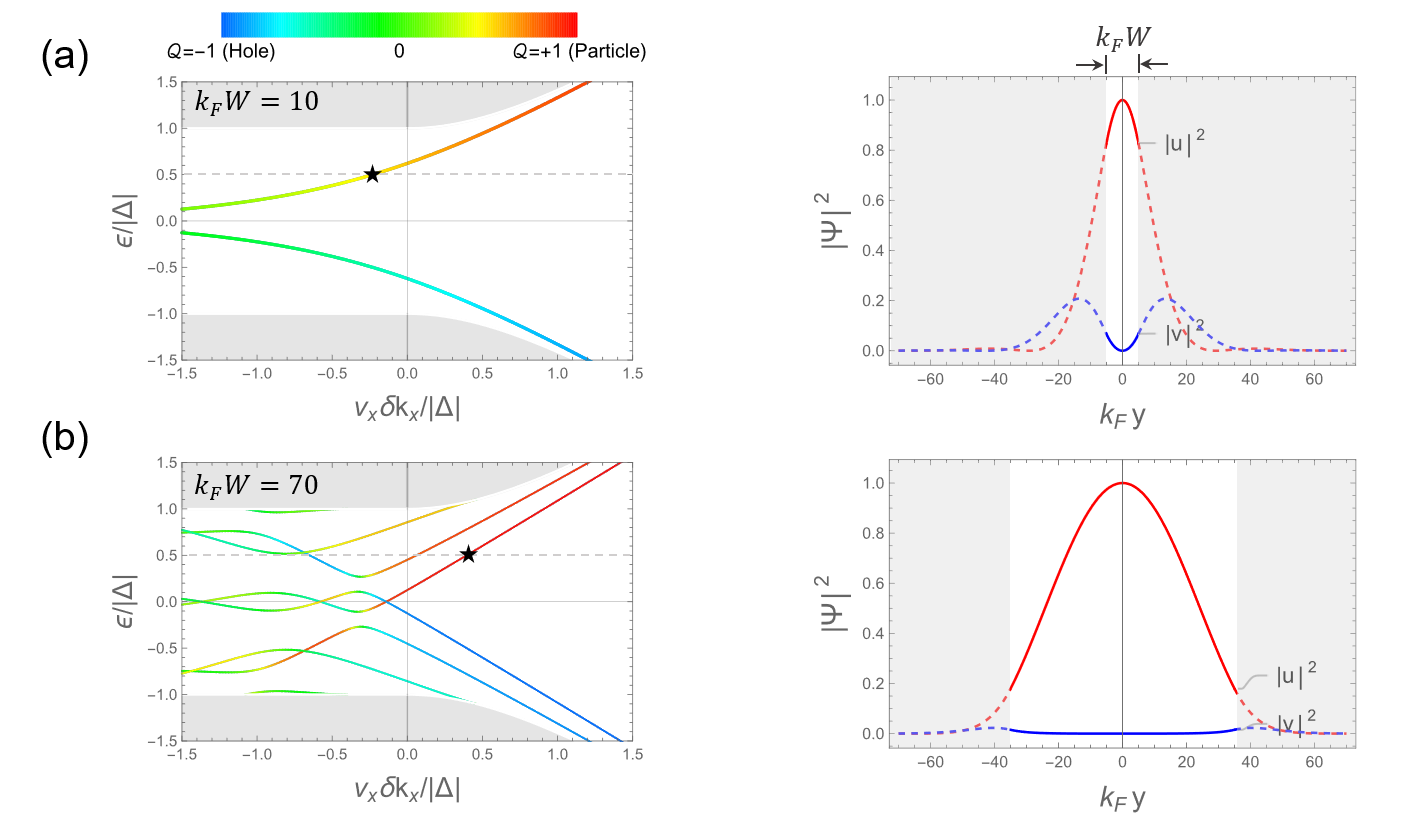}\centering
  \caption{\small{Left: BdG spectrum in an SNS $\pi$-junction of width (a) $W=10k^{-1}_F$ and (b) $W=70k^{-1}_F$. Here we have taken $ak_F^2/\abs{\Delta}=100$}, where $k_F$ is the Fermi wavevector. Each state in the spectrum is colored according to its $\mathcal{Q}$-value, which reflects its particle-hole nature. Shaded regions represent the bulk states. Following a particular ABS (of a fixed $\varepsilon$) as $W$ increases, we see that its $\mathcal{Q}$-value approaches $\pm 1$. Right: wavefunction $\Psi(y) = (u,v)^T$ of a representative state marked by a star in the spectrum. The superconducting region, with $\abs{y}>W/2$, is shaded. The chosen state has $\varepsilon=0.5 \abs{\Delta}$. Comparison between (a) and (b) demonstrates the fate of this ABS as it transits from the narrow junction into the normal lead.}
  \label{supp:fig:finiteW}
\end{figure}

The ABS is determined by matching two boundary conditions at the SN interface: $\Psi_{\text I}(W/2) = \Psi_\text{II}(W/2)$ and $\partial_y \Psi_{\text I}(W/2) = \partial_y \Psi_\text{II}(W/2)$. These can be summarized as $\mathcal{B}(\Gamma_e, \Gamma_h, \gamma^+, \gamma^-)^T=0$, with
\begin{equation}
     \mathcal{B}= \begin{pmatrix}
    \cos{\frac{k_e W}{2}} & 0 & -1 & -1 \\
    0 & \sin{\frac{k_h W}{2}} & -ie^{\sgn(a)i\eta} & -ie^{-\sgn(a)i\eta} \\
   -k_e\sin{\frac{k_e W}{2}} &0 & \kappa^+ &\kappa^- \\
    0 & k_h\cos{\frac{k_h W}{2}} & i\kappa^+ e^{\sgn(a)i\eta} & i\kappa^- e^{-\sgn(a)i\eta}
    \end{pmatrix}.
\end{equation}
A non-trivial solution requires $\det \mathcal{B}=0$. Solving this gives us all the allowed energies at a given $\delta k_x$, from which we obtain the dispersion of all ABSs inside the junction with width $W$. The corresponding wavefunctions can be read-off from the null-space of $\mathcal{B}$. In addition to the spectrum, we are also interested in the particle-hole character of the ABS, as we want to determine the fate of an ABS as it propagates into the lead (i.e. as $W$ increases). Thus, for the wavefunction $\Psi(y) = (u,v)^T$, let us assign the following quantity to characterize the particle-hole nature of the state:
\begin{equation}
    \mathcal{Q} = \frac{\int^\infty_{-\infty} dy\; (\abs{u}^2 - \abs{v}^2)}{\int^\infty_{-\infty} dy\; (\abs{u}^2 +\abs{v}^2)}.
\end{equation}
As such, a definitive particle-state has $\mathcal{Q}=+1$, while a definitive hole-state has $\mathcal{Q}=-1$. 

Figure \ref{supp:fig:finiteW} (a,b) (left panels) show the BdG spectra for two junction widths $W$, where each ABS is displayed in a color reflecting its $\mathcal{Q}$-value. For each $W$, the wavefunction of a representative state (chosen at $\varepsilon=0.5 \abs{\Delta}$) is plotted in the right panel. We have illustrated here the case with $a>0$, so the normal dispersion is electron-like. An ABS with $\varepsilon>0$ then converts definitively into an electron as it propagates from a narrow junction ($W \ll \xi$) into the lead ($W \gg \xi$). On the other hand, if $a<0$, it would be the ABS with $\varepsilon<0$ that acquires $\mathcal{Q}\rightarrow +1$ as $W \rightarrow \infty$. The physical states thus belong to a hole-like band, and hence the ABS excitation would convert definitively into a hole inside the lead. Notice that we have arrived at the same conclusion as we did in the main text, while there we considered taking $\abs{\Delta} \rightarrow 0$ instead of $W \rightarrow \infty$. 

Last but not least, let us remark on the length scales involved in this problem. In a sense, there are two superconducting coherence lengths. Away from the critical point where the dispersion is linear, i.e. $E(k_y) = v_F k_y$, the level-spacing is roughly $v_F/W$. Close to the critical point where the dispersion is quadratic, i.e. $E(k_y) = ak_y^2$, the spacing becomes $\abs{a}/W^2$. The condition for a \textit{narrow} junction, which hosts one branch of ABS for each Fermi surface critical point, is that $\min(v_F/W, \abs{a}/W^2) \gtrsim \abs{\Delta}$. Denoting the usual coherence length as $\xi=v_F/\abs{\Delta}$, and the one associated to the critical point as $\xi_c=\sqrt{\abs{a}/\abs{\Delta}}$, we see that $\xi_c \ll \xi$ for $E_F \gg \abs{\Delta}$. This feature is reflected in Fig. \ref{supp:fig:finiteW}(b), where we see that the level-spacing is larger for $v_x\delta k_x/\abs{\Delta}\lesssim -1$, compared to the spacing for $v_x\delta k_x/\abs{\Delta} \approx 0$. Our proposal requires a narrow junction that satisfies $W\lesssim \xi_c$. 

\subsection{Landauer formalism for the proposed setup}\label{supp:LB}
\setcounter{equation}{0}
\setcounter{figure}{0}    
Here we derive an expression for the net current $I_p$ flowing into lead $p$ in our proposed setup. In particular, we obtain Eq. (10) in the main text. The net current $I_p$ in lead $p$ is contributed by the following scattering processes. For an incident electron-excitation ($E>0$) in the $n$-th transport mode moving from lead $p$ into the junction, it can be either normally-reflected back into lead $p$ as an electron (with probability $r^N_{n,p}(E)$), or Andreev reflected back into lead $p$ as a hole (with probability $r^A_{n,p}(E)$), or \textit{crossed} Andreev reflected into lead $q \neq p$ as a hole (with probability $t^A_{n,qp}(E)$). As for an incident hole-excitation ($E<0$), there are three similar scattering processes, and the corresponding scattering probabilities are denoted in the same way, but with $E<0$. As such, the net current flowing into lead $p$ can be expressed as
\begin{equation}\label{supp:eq:Ip}
\begin{split}
I_p = -\frac{2e}{h}\sum_n\Big\{\int_{0}^\infty dE\;\big[(1-r^N_{n,p}+r^A_{n,p})f_p(E) - \sum_{q\neq p}(t^N_{n,pq}-t^A_{n,pq})f_q(E)\big]\\
 - \int_{-\infty}^0 dE\;\big[(1-r^N_{n,p}+r^A_{n,p})\bar{f}_p(E)  - \sum_{q\neq p}(t^N_{n,pq}-t^A_{n,pq})\bar{f}_q(E)\big] \Big\}.
\end{split}
\end{equation}
Here $f_p(E) = f(E-eV_p)$, with $f(E)=(1+e^{\beta E})^{-1}$, and $\bar{f}_p(E)=1-f_p(E)$, which are the respective Fermi distributions for electrons and holes in lead $p$. The factor of 2 upfront accounts for spin-degeneracy. Unitarity demands
\begin{equation}\label{supp:eq:unitarity}
r^N_{n,p} + r^A_{n,p} +\sum_{q\neq p} (t^N_{n,qp}+t^A_{n,qp}) =1.
\end{equation}
Notice that we assume $t^A_{n,qp}=0$, i.e. crossed Andreev reflection is suppressed. This can be achieved by a \textit{long} Josephson junction, with length $L$ much larger than the coherence length $\xi$ of the superconductor (see Fig. 1(a) in the main text). Eliminating $r^N_{n,p}$, we obtain
\begin{equation}
\begin{split}
I_p = -\frac{2e}{h}\sum_n\Big\{\int_{0}^\infty dE\;\big[\sum_{q\neq p}\big(t^N_{n,qp}f_p(E)-t^N_{n,pq}f_q(E)\big)+2r^A_{n,p}f_p(E)\big]\\
 - \int_{-\infty}^0 dE\;\big[\sum_{q\neq p}\big(t^N_{n,qp}\bar{f}_p(E)-t^N_{n,pq}\bar{f}_q(E)\big)+2r^A_{n,p}\bar{f}_p(E)\big] \Big\}.
\end{split}
\end{equation}
Let us define the normal transmission functions, as well as the Andreev reflection functions, for electrons/holes as follows,
\begin{equation}
    T^{e/h}_{qp}(E) = \begin{cases}
    \sum_n t^N_{n,qp} (E), &E \gtrless 0\\
     0, &E \lessgtr 0
    \end{cases}\;\;\;\;\;\;\text{and}\;\;\;\;\;\;
    R^{e/h}_{p}(E) = \begin{cases}
    \sum_n r^A_{n,p} (E), &E \gtrless 0\\
     0, &E \lessgtr 0
    \end{cases}.
\end{equation}
We then obtain $I_p = I^N_p +I^A_p$, with
\begin{subequations}
\begin{align}
 I^N_p  &= -\frac{2e}{h}\sum_{q \neq p} \int_{-\infty}^{+\infty}[T^e_{qp}(E) f_p(E) - T^e_{pq}(E) f_q(E)-T^h_{qp}(E) \bar{f}_p(E) + T^h_{pq}(E) \bar{f}_q(E)]dE, \\
 I^A_p & = -\frac{4e}{h} \int_{-\infty}^\infty \big[R^e_{p}(E) f_p(E) - R^h_{p}(E)\bar{f}_p(E) \big]dE.
 \end{align}
\end{subequations}

Now getting back to the measurement protocol specified in the main text. We are interested in measuring $I_2$ as a function of $V_1$, by holding $V_2=0$ fixed. As lead 2 is \textit{not} biased from the superconductor, $I^A_2=0$. We also have $I_2(V_1=0)=0$, as there should be no flow of current at equilibrium. Altogether, 
\begin{equation}
\begin{split}
    I_2(V_1) = I^N_2(V_1) -  I^N_2(0) &= \frac{2e}{h} \int_{-\infty}^{+\infty} [T^e_{21}(E) + T^h_{21}(E)]\cdot[f_1(E)-f_2(E)]\;dE \\
    &= \frac{2e}{h} \int_{-\infty}^{+\infty} [T^e_{21}(E) + T^h_{21}(E)]\cdot[f(E-eV_1)-f(E)]\;dE.
\end{split}
\end{equation}
This is Eq. (10) in the main text. 

\subsection{Dispersive Andreev states with Rashba SOC}\label{supp:SOC}
\setcounter{equation}{0}
\setcounter{figure}{0}   

\begin{figure}[h]
  \includegraphics[width=0.9\columnwidth ]{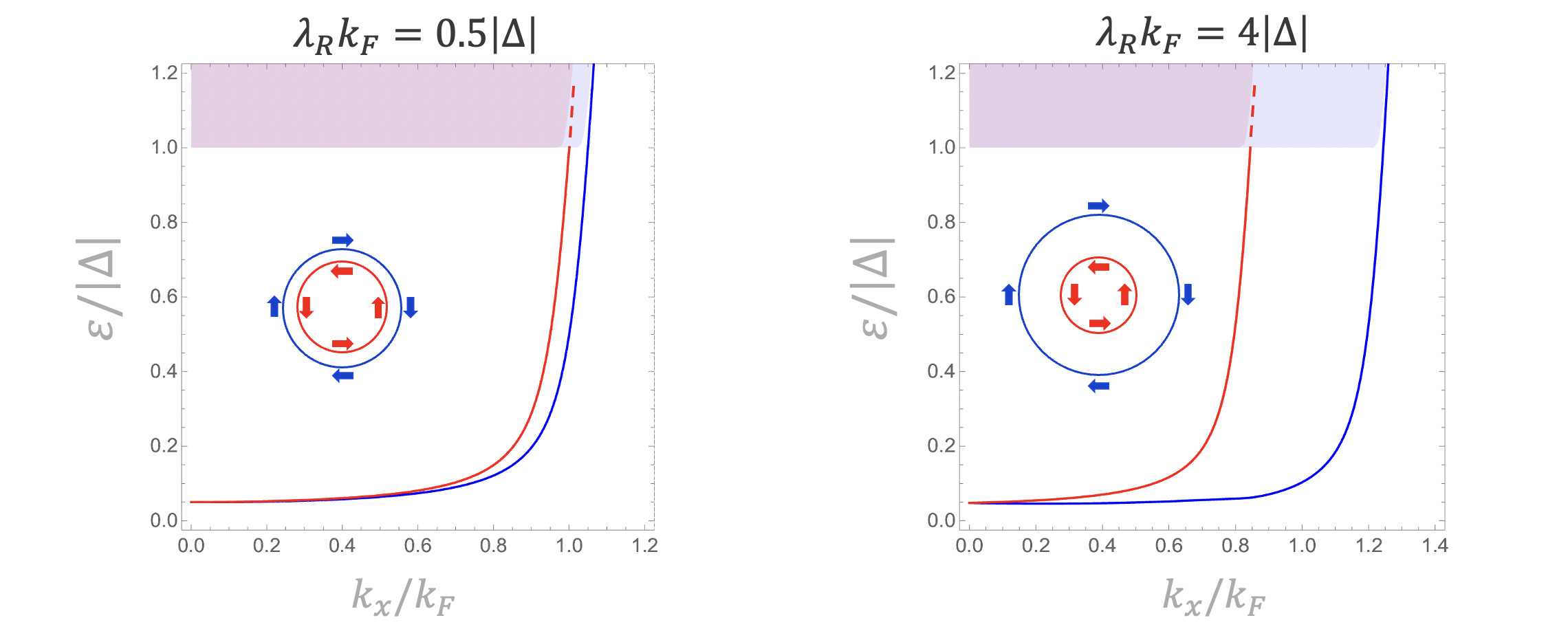}\centering
  \caption{\small{Dispersion of Andreev bound states for the two Rashba-split Fermi pockets, labeled as red and blue curves respectively. Here we have chosen $ak_F^2=E_F=10\abs{\Delta}$, where $k_F$ is the Fermi wavevector for the unsplit Fermi surfaces with $\lambda_R=0$. Only the positive energy spectrum is shown. The shaded region corresponds to the continuum of bulk states. While the inner branch of ABS (shown in red) only exists for $\abs{\varepsilon}<\abs{\Delta}$, the outer branch remains a valid bound state for $\abs{\varepsilon}>\abs{\Delta}$. As $\abs{\Delta}\rightarrow 0$, both ABSs merge with the bulk continuum and acquire the electron/hole nature as dictated by $\sgn(a)$.}}
  \label{supp:fig:RashbaABS}
\end{figure}

Here we solve for the dispersive ABSs in the presence of Rashba SOC. We confirm the argument presented in the main text, which suggests that there is one branch of electron/hole-like ABS for each electron/hole-like Rashba-split Fermi pocket, see Fig. \ref{supp:fig:RashbaABS}. Similar to the case without SOC, the ABS is an approximate zero-mode ($\varepsilon \approx \abs{\Delta^2/(2E_F)}$) inside the Fermi pocket, and becomes strongly dispersive past a Fermi surface critical point. Our proposal for measuring Fermi sea topology thus applies to two-dimensional electron gas (2DEG) with SOC, leading to a quantized conductance that probes $\chi^{R}_F+\chi^{L}_F$, where $R/L$ refers to the chirality of spin-momentum locking: $\sgn[({\bf k}\times {\bf \sigma})\cdot \hat{z}] =\pm 1$.

Let us consider the following BdG Hamiltonian:
\begin{equation}\label{supp:eq:HBdG}
    H_{\text{BdG}}=[ a(k_x^2-\partial^2_y)-E_F+\lambda_R(k_x \sigma_y +i\partial_y \sigma_x) ]  \tau_z  + \text{sgn}(y) \abs{\Delta} \tau_y.
\end{equation}
As in the main text, we consider the extreme narrow-junction limit ($W=0$) and model the lead by taking $\abs{\Delta}\rightarrow 0$. Here, $\tau$'s are the Pauli matrices acting on the particle-hole space, and $\sigma$'s are the Pauli matrices acting on the physical spin space. The strength of Rashba SOC is characterized by $\lambda_R \geq 0$. The effective mass $m$ of the 2DEG is related to the model parameter $a$ by $a=\hbar^2/(2m)$, thus an electron/hole-like Fermi pocket has sgn$(a)=\pm 1$. In addition to the chiral symmetry $\Pi = \tau_x$, with $\{H_{\text{BdG}},\Pi\}=0$, we also have a mirror-$y$ symmetry $\mathcal{M}_y = M_y \sigma_y \tau_z$, where $M_y$ flips the real-space coordinate $y \mapsto -y$. It is straightforward to check that $[H_{\text{BdG}}, \mathcal{M}_y] =0$. Moreover $\{\mathcal{M}_y,\Pi\}=0$, hence particle-hole partners (at the same $k_x$ and with energy $\pm \varepsilon$, as related by $\Pi$) acquire opposite mirror eigenvalues.

Next, we construct the bound-state ansatz $\Psi_{y>0}$ that diagonalizes $H_{\text{BdG}}$ in the upper-half plane. Inside the superconductor, the bound-state wavefunction should be a linear combination of decaying modes $e^{ik_y y}$, with $\Im(k_y)>0$. Each mode $e^{ik_y y}$ is associated with a spinor $d^\sigma_s(k_y)$ that diagonalizes the SOC term:
\begin{equation}
    (k_x\sigma_y-k_y\sigma_x)\;d^\sigma_s(k_y) = s\sqrt{k_x^2+k_y^2}\;d^\sigma_s(k_y) \quad \implies \quad d^\sigma_s(k_y)= \Big(\;1\;,\; s\cdot\frac{ik_x-k_y}{\sqrt{k_x^2+k_y^2}}\;\Big)^T, 
\end{equation}
with $s=\pm 1$, and $\sqrt{k_x^2+k_y^2}$ is defined to have a positive real part.
The complex momentum $k_y$ then obeys the following characteristic equation for the BdG eigen-energy $\varepsilon$:
\begin{equation}
\begin{split}
    &\Big[\;a(k_x^2+k_y^2)-E_F+s\lambda_R\sqrt{k_x^2+k_y^2}\;\Big]^2 = \varepsilon^2-\abs{\Delta}^2 \\ 
    \implies\quad &\sqrt{k_x^2+k_y^2} = \frac{1}{2\abs{a}}\Big[-s\cdot\sgn(a)\lambda_R+\sqrt{\lambda^2_R+4aE_F\pm4ai\sqrt{\abs{\Delta}^2-\varepsilon^2}}\;\Big]
\end{split}
\end{equation}
This gives four solutions to $k_y$ that satisfy $\Im(k_y)>0$. For $s\cdot\sgn(a)=+1$, we have two solutions $k_{+,1}$ and $k_{+,2}$:
\begin{equation}
    k_{+,1/2} = \pm \sqrt{-k_x^2+\frac{1}{4a^2}\Big(\lambda_R-\sqrt{\lambda^2_R+4aE_F\pm4\abs{a}i\sqrt{\abs{\Delta}^2-\varepsilon^2}}\Big)^2}.
\end{equation}
Here, the spinor that diagonalizes the physical spin space is $d^\sigma_{\sgn(a)}$, while the corresponding spinor that diagonalizes the particle-hole space is $(1, ie^{\mp \sgn(a) i\eta})^T$, with $e^{i\eta} := \varepsilon/\abs{\Delta}+i\sqrt{1-(\varepsilon/\abs{\Delta})^2}$. For $s\cdot\sgn(a)=-1$, we have two solutions $k_{-,1/2}$:
\begin{equation}
    k_{-,1/2} = \pm \sqrt{-k_x^2+\frac{1}{4a^2}\Big(\lambda_R+\sqrt{\lambda^2_R+4aE_F\pm4\abs{a}i\sqrt{\abs{\Delta}^2-\varepsilon^2}}\Big)^2}.
\end{equation}
The spinor that diagonalizes the physical spin subspace is $d^\sigma_{-\sgn(a)}$, and the corresponding spinor that diagonalizes the particle-hole subspace is $(1, ie^{\mp \sgn(a) i\eta})^T$. Altogether, we can construct a generic ansatz to the uniform BdG equation in the upper-half plane:
\begin{equation}
\begin{split}
    \Psi_{y>0}  =\quad & (\gamma_{+,1} e^{ik_{+,1}y} \cdot d^\sigma_{\sgn(a)} + \gamma_{-,1}e^{ik_{-,1}y} \cdot d^\sigma_{-\sgn(a)}) \otimes \begin{pmatrix}
1 \\ ie^{-\text{sgn}(a)i\eta}
\end{pmatrix}_\tau\\
+& (\gamma_{+,2} e^{ik_{+,2}y} \cdot d^\sigma_{\sgn(a)} + \gamma_{-,2}e^{ik_{-,2}y} \cdot d^\sigma_{-\sgn(a)})\otimes \begin{pmatrix}
1 \\ ie^{\text{sgn}(a)i\eta}
\end{pmatrix}_\tau\;.
\end{split}
\end{equation}
The subscript $\tau$ indicates spinors in the particle-hole subspace, and $\gamma$'s are coefficients to be determined by the boundary conditions.


We now construct the bound-state wavefunction in the lower-half plane, $\Psi_{y<0}$, and match boundary conditions to find $\varepsilon(k_x)$. Using the mirror symmetry $\mathcal{M}_y = M_y \sigma_y \tau_z$, we shall assume $\Psi_{y<0} = \mathcal{M}_y \Psi_{y>0}$. The chiral symmetry $\Pi$, which anti-commutes with both $H_{\text{BdG}}$ and $\mathcal{M}_y$, allows us to obtain another solution with energy $-\varepsilon$ and $\mathcal{M}_y=-1$, thus forming the full BdG spectrum. The quadratic BdG equation requires two boundary conditions, $\Psi(0^+)=\Psi(0^-)$ and $\partial_y\Psi(0^+)=\partial_y\Psi(0^-)$, which can be summarized as $\mathcal{B} (\gamma_{+,1}, \gamma_{-,1}, \gamma_{+,2}, \gamma_{-,2})^T =0$, with
\begin{equation}
    \mathcal{B}= \begin{pmatrix}
    f(a,k_{+,1}) & f(-a,k_{-,1}) & f(a,k_{+,2}) & f(-a,k_{-,2}) \\
    f(-a,k_{+,1}) & f(a,k_{-,1}) & e^{2\sgn(a)i\eta}f(-a,k_{+,2}) & e^{2\sgn(a)i\eta}f(a,k_{-,2}) \\
    f(-a,k_{+,1})\cdot k_{+,1} & f(a,k_{-,1})\cdot k_{-,1} & f(-a,k_{+,2})\cdot k_{+,2} & f(a,k_{-,2})\cdot k_{-,2} \\
    f(a,k_{+,1})\cdot k_{+,1} & f(-a,k_{-,1})\cdot k_{-,1} & e^{2\sgn(a)i\eta}f(a,k_{+,2})\cdot k_{+,2} & e^{2\sgn(a)i\eta}f(-a,k_{-,2})\cdot k_{-,2}
    \end{pmatrix}
\end{equation}
and $f(a,k_y) = 1+i\sgn(a)\frac{ik_x-k_y}{\sqrt{k_x^2+k_y^2}}$. From $\det \mathcal{B} = 0$, we can self-consistently solve for the dispersion of ABS. Resorting to numerical solutions, we obtain two branches of ABSs, one for each Rashba-split Fermi pocket. The results are shown in Fig. \ref{supp:fig:RashbaABS}.

\end{document}